\documentclass[12pt]{article}
\usepackage{amssymb,amsmath,epsfig}
\usepackage{graphicx}
\allowdisplaybreaks
\begin{document}

\title{\bf Impact of Charge on the Complexity of Static Sphere in $f(R,\textbf{T}^{2})$ Gravity}

\author{M. Sharif \thanks{msharif.math@pu.edu.pk} and Ayesha Anjum
\thanks{ayeshaanjum283@gmail.com}\\
Department of Mathematics, University of the Punjab,\\
Quaid-e-Azam Campus, Lahore-54590, Pakistan.}

\date{}

\maketitle
\begin{abstract}
This paper investigates the complexity of a charged static sphere
filled with anisotropic matter in the background of energy-momentum
squared gravity. For this purpose, we evaluate the modified field
and conservation equations to determine the structure of celestial
system. The mass function is calculated through Misner-Sharp as well
as Tolman mass definitions. The complexity of a self-gravitating
system depends on different factors such as anisotropic pressure,
electromagnetic field, energy density inhomogeneity, etc. We
formulate the structure scalars by the orthogonal decomposition of
the Riemann tensor to develop a complexity factor containing all
vital features of the stellar structure. The vanishing complexity
condition is achieved by setting the complexity factor equal to
zero. Finally, we construct two static solutions by utilizing the
energy density of Gokhroo-Mehra solution as well as the polytropic
equation of state along with the zero complexity condition. It is
found that electromagnetic field decreases the complexity of stellar
structure.
\end{abstract}

\section{Introduction}

Several astronomical observations from various experiments
(Two-degree Field Galaxy Redshift, Survey Sloan Digital Sky Survey,
Large Synoptic Survey Telescope) prove that the large scale
structures such as galaxies and stars give vital information about
the origin and evolution of the vast universe. Therefore, the study
of these compact structures play a significant role in understanding
the mechanism and origin of the cosmos. These self-gravitating
structures are complex in nature as they comprise of organized and
inter-linked components that collaborate in different ways. A slight
perturbation in the complicated system may cause principal changes
within the physical attributes of the stellar structure. To analyze
the complicated nature and evolution of the astrophysical structure,
it is mandatory to define a complexity factor that connects the
essential physical parameters. Moreover, an effective complexity
factor must measure how the internal and external perturbations
affect the stability and evolution of stellar structures.

Numerous researchers have put a lot of efforts in determining an
appropriate definition of complexity in different divisions of
science \cite{1}. However, a standard definition that is applicable
in all fields has not been accomplished. In the earlier definitions
of complexity, the idea of data and entropy were taken into account.
However, these definitions were unable to precisely evaluate the
complexity of the basic physical models: ideal gas and perfect
crystal. In order to describe the symmetric distribution of
particles in the perfect crystal, minimal data is sufficient. On the
other hand, in case of ideal gas, maximal information is needed to
indicate any of its likely state. Although these two states display
a contrasting behavior, both are allotted the least complexity.
Therefore, the definition of complexity must include factors other
than entropy and information.

Lopez-Ruiz et al. \cite{3} extended the idea of complexity by
examining the concept of disequilibrium which is the difference
between various probabilistic states and the equiprobable
configuration of the structure. Researchers have replaced the
probability distribution (used in defining the disequilibrium and
information) by the energy density of the system to compute the
complexity of self-gravitating structures like neutron stars and
white dwarfs \cite{4}. Dense stellar systems have particles that are
compactly arranged in their interior. Consequently, less radial
pressure is generated as compared to the tangential pressure which
produces anisotropy in pressure. Thus, the anisotropy is an
essential factor in determining the stability of a compact
structure. As the idea of complexity suggested by Lopez-Ruiz and his
collaborators dealt with energy density only while other state
variables such as pressure and anisotropy were neglected, therefore,
it does not define an effective criterion to measure complexity.

Recently, Herrera \cite{5} established a new factor of complexity
for static spherically symmetric structure in the background of
general relativity (GR). The structure of the spherical system was
examined in terms of state determinants such as anisotropic
pressure, energy density, etc. The basic assumption was that a
static matter distribution with homogeneous density and isotropic
pressure constituted the simplest system. Consequently, the
complexity factor for such a distribution is zero. Herrera utilized
the Tolman mass to relate inhomogeneous energy density and pressure
anisotropy through a single relation. He developed a complexity
factor for self-gravitating anisotropic source by utilizing Bel's
technique for orthogonal decomposition of the Riemann tensor. Sharif
and Butt \cite{6} examined the complexity of the static cylindrical
self-gravitating system by employing Herrera's approach. Herrera et
al. \cite{7} also discussed the complexity of a dynamical
spherically symmetric fluid distribution following a homologous
pattern of evolution. This definition has been extended to axially
symmetric static case by formulating three complexity factors
\cite{8}. Recently, Herrera et al. \cite{9} explored a
quasi-homologous self-gravitating structure satisfying the zero
complexity condition.

The electromagnetic field is an essential factor in the evolution
and stability of self-gravitating objects as it overcomes the
attractive gravitational force. In astrophysics, the
Reissner-Nordstr\"{o}m metric was the first static spherically
symmetric charged solution to the Einstein-Maxwell field equations.
Rosseland and Eddington \cite{10} studied the characteristics of
charged fluid in the interior of self-gravitating spherically
symmetric structures. Ray et al. \cite{11} investigated the effect
of electromagnetic field on compact stellar structure and deduced
that the local impact of forces exerted on single charged particle
created an imbalance of forces which ultimately generated a charged
black hole. Sharif and Bhatti \cite{12} investigated the stability
of charged sphere filled with viscous dissipative matter and
discussed the stability regions for collapsing systems through the
adiabatic index. The stability of spherically symmetric anisotropic
polytrope under the impact of electromagnetic field was checked by
Sharif and Sadiq \cite{13}. Sharif and Butt formulated the
structures scalars by adopting Herrera's technique and calculated
the complexity factors for charged spherical \cite{14} as well as
cylindrical \cite{15} self-gravitating systems. They concluded that
the electromagnetic field decreases the complexity of stellar
structures.

According to cosmic observations from numerous surveys (Type Ia
Supernovae \cite{16}, Cosmic Microwave Background Radiation
\cite{17}, Large Scale Structure \cite{18}, Baryon Acoustic
Oscillations \cite{19} and Sloan Digital Sky Survey \cite{20}), our
universe is undergoing accelerated expansion. This accelerated
expansion is supposed to be the result of some mysterious force
known as dark energy that has large negative pressure. Two different
methods have been employed by researchers to determine the cause of
cosmic expansion. First approach requires the modification of the
energy-momentum tensor while the other way is to alter the geometric
part in the Einstein-Hilbert action which leads to modified theories
of relativity.

In GR, $\Lambda$ Cold Dark Matter model is used to successfully
describe the evolution of the cosmos. However, it has some issues
namely fine-tuning and coincidence problems. Thus, modified theories
like $f(R),~f(R, T)$, etc. ($R$ denotes the Ricci scalar and $T$ is
the trace of energy-momentum tensor $T_{\mu\nu}$) have gained the
attention of researchers to deal with issues related to cosmic
acceleration. The $f(R)$ theory is achieved by replacing $R$ with
the generic function $f(R)$ in the Einstein-Hilbert action
\cite{21}. Harko et al. \cite{22} proposed the $f(R,T)$ theory (an
extension of $f(R)$ theory) by considering a gravitational
Lagrangian density in terms of $R$ and $T$. The curvature-matter
coupling models in $f(R,T)$ theory are valuable for describing the
late-time cosmic acceleration as well as the interconnection of dark
energy and dark matter \cite{22*}. A systematic review of some
standard issues and also the latest developments of modified gravity
in cosmology is given in \cite{22a}. The measure of complexity,
introduced by Herrera, has also been developed in the context of
these modified theories. Abbas and Nazar computed the complexity
factor in $f(R)$ gravity for static \cite{23} as well as dynamical
\cite{24} fluid distribution. Abbas and Ahmad \cite{25} analyzed the
complexity for a class of compact stars in $f(R,T)$ scenario.
Herrera's technique to formulate complexity factor has also been
applied in other modified theories of gravity \cite{27}. Several
other works studying compact objects in modified gravity can be seen
in \cite{27a}.

Recently, Katrici and Kavuk \cite{31} proposed a new generalization
of GR by defining a specific coupling between matter and gravity
through a term proportional to $T^{\mu\nu}T_{\mu\nu}$. This theory
is referred to as energy-momentum squared gravity (EMSG) or
$f(R,\textbf{T}^{2})$ theory with
$\textbf{T}^{2}=T^{\mu\nu}T_{\mu\nu}$. The predictions of GR about
singularities at high energy levels (such as big bang singularity)
are no longer valid due to expected quantum effects. In this
respect, EMSG is considered as a favorable framework because it
resolves the big bang singularity by supporting regular bounce with
finite maximum energy density and least scale factor in the
beginning of the cosmos. The conservation law does not hold in EMSG
due to interaction between matter and curvature which indicates the
presence of some extra force. Consequently, the path of the test
particle differs from the standard geodesic path. Various
astrophysical and cosmological structures have been studied in
$f(R,\textbf{T}^{2})$ theory.

Roshan and Shojai \cite{34} obtained the exact solution of EMSG
field equations and determined the possibility of bounce at an early
time by applying $f(R,\textbf{T}^{2})$ theory to homogeneous and
isotropic spacetime. Broad and Barrow \cite{35} studied exact
solutions representing the isotropic universe for different forms of
$f(R,\textbf{T}^{2})$ and discussed their behavior with reference to
various physical parameters. Some astrophysical objects such as
neutron stars were examined in the background of EMSG with
$f(R,\textbf{T}^{2})=R+\zeta T^{\mu\nu}T_{\mu\nu}$, $\zeta$ being a
constant \cite{32}. Bahamonde et al. \cite{36} examined the dynamics
of two different models to explain the current accelerated cosmic
expansion. Sharif and Gul \cite{37} explored the structure of cosmic
objects through the Noether symmetry approach. They also studied the
dynamics of cylindrical collapse with dissipative matter in the
presence of charge and deduced that the dissipative matter,
electromagnetic field and modified terms reduce the collapse rate
\cite{38}.

The objective of this article is to develop the vanishing complexity
condition for a static spherical distribution in the presence of
charge within $f(R,\textbf{T}^{2})$ background. The layout of the
paper is as follows. In the next section, we formulate the EMSG
field equations for an anisotropic matter distribution. We discuss
some physical attributes of matter distribution in section
\textbf{3}. In section \textbf{4}, structure scalars are constructed
by decomposing the curvature tensor with the help of four-velocity.
In section \textbf{5}, we formulate the zero complexity condition to
produce solutions of modified field equations corresponding to an
assumed energy density as well as polytropic equation of state.
Finally, in section \textbf{6} we summarize the main results.

\section{Field Equations}

In this section, we will describe some physical variables related to
charged spherical stellar structure and obtain the corresponding
field equations. The modified Einstein-Hilbert action in
$f(R,\textbf{T}^{2})$ gravity is given as \cite{31}
\begin{equation}\label{1}
S=\int(\mathcal{L}_m+\mathcal{L}_e)\sqrt{-g}d^{4}x+\int\frac{f(R,\textbf{T}^{2})}{2\kappa^{2}}\sqrt{-g}d^{4}x,
\end{equation}
where $g,~\mathcal{L}_m,~\mathcal{L}_e$ and $\kappa$ are the
determinant of the metric tensor ($g_{\mu\nu}$), matter Lagrangian,
Lagrangian for the electromagnetic field and coupling constant,
respectively. Here $\mathcal{L}_e$ has the form
\begin{equation*}
\mathcal{L}_e=-\frac{1}{16\pi}\mathit{F}^{\mu\nu}\mathit{F}_{\mu\nu},
\end{equation*}
where the Maxwell field tensor $\mathit{F}_{\mu\nu}$ is defined as
$\mathit{F}_{\mu\nu}=\phi_{\nu,\mu}-\phi_{\mu,\nu}$ and
$\phi_a=\phi\delta^{0}_a$ with $\phi$ as the scalar field potential.
The EMSG field equations obtained by varying Eq.(\ref{1}) are
\begin{equation}\label{2}
R_{\mu\nu}f_R+g_{\mu\nu}\Box f_R-\nabla_{\mu}\nabla_{\nu}f_R-
\frac{1}{2}g_{\mu\nu}f=\kappa^{2}T_{\mu\nu}+E_{\mu\nu}-\Theta_{\mu\nu}f_{\textbf{T}^{2}},
\end{equation}
where $R_{\mu\nu}$ denotes the Ricci tensor. Also,
$\Box=\nabla^{\mu}\nabla_\mu$, $f_R=\frac{\partial f}{\partial R}$
and $f_{\textbf{T}^{2}}=\frac{\partial f}{\partial \textbf{T}^{2}}$.
Here ${E}_{\mu\nu}$ is the electromagnetic field tensor and
$\Theta_{\mu\nu}$ is given as
\begin{eqnarray}\nonumber
\Theta_{\mu\nu}=\frac{\delta \textbf{T}^{2}}{\delta
g^{\mu\nu}}=\frac{\delta(T^{\mu\nu}T_{\mu\nu})}{\delta
g^{\mu\nu}}&=&-2\mathcal{L}_m(T_{\mu\nu}-\frac{1}{2}g_{\mu\nu}T)
-4\frac{\partial^{2}\mathcal{L}_m}{\partial g^{\mu\nu}\partial
g^{\alpha\beta}}T^{\alpha\beta}\\\label{3} &-&T T_{\mu\nu}+2
T^{\alpha}_\mu T_{\nu\alpha}.
\end{eqnarray}

The energy-momentum tensor related to the anisotropic fluid
distribution is expressed as
\begin{equation}\label{4}
T^{\mu}_\nu=\rho\emph{v}^{\mu}\emph{v}_\nu-\emph{p}\emph{h}^{\mu}_\nu+\Pi^{\mu}_\nu.
\end{equation}
Here $\emph{v}_\mu,~\emph{p},~\rho $ and $\Pi_{\mu\nu}$ denote the
four-velocity, pressure, energy density and anisotropic tensor,
respectively. These terms are defined as
\begin{eqnarray*}
\Pi^{\mu}_\nu=\frac{\Pi}{3}(3\emph{s}^{\mu}\emph{s}_\nu+\emph{h}^{\mu}_\nu),\quad
\Pi=\emph{p}_r-\emph{p}_\bot,\quad\emph{h}^{\mu}_\nu=\delta^{\mu}_\nu-\emph{v}^{\mu}\emph{v}_\nu,~\emph{p}=\frac{1}{3}(\emph{p}_r
+2\emph{p}_\bot),
\end{eqnarray*}
where $\emph{p}_r$ and $\emph{p}_\bot$ are the radial and tangential
pressures of anisotropic fluid, respectively. As matter Lagrangian
has no specific definition, therefore, different matter Lagrangians
produce different forms of the field equations. More widely used
forms of matter Lagrangian are $\mathcal{L}_m=-\emph{p}$ and
$\mathcal{L}_m=\rho$. These choices do not pose any problem in GR.
However, for non-minimal coupling case, different forms of matter
Lagrangian correspond to distinct results \cite{39}. Thus, for our
convenience, we consider $\mathcal{L}_m=\rho$ and $\kappa=1$ which
yields \cite{40}
\begin{eqnarray}\nonumber
\Theta_{\mu\nu}&=&-2\rho(T_{\mu\nu}-\frac{1}{2}g_{\mu\nu}T)
-TT_{\mu\nu}+2 T^{\alpha}_\mu T_{\nu\alpha},
\\\label{5} \mathbb{G}_{\mu\nu}&=&R_{\mu\nu}-\frac{1}{2}R
g_{\mu\nu}=\frac{1}{\kappa^{2}f_R}(T^{(C)}_{\mu\nu}+T_{\mu\nu}+\emph{E}_{\mu\nu})=T^{(D)}_{\mu\nu},
\end{eqnarray}
where $\mathbb{G}_{\mu\nu}$ is the Einstein tensor and
$T^{(C)}_{\mu\nu}$ are the modified terms of $f(R,\textbf{T}^{2})$
gravity (also called correction terms) takes the form
\begin{eqnarray}\nonumber
T^{(C)}_{\mu\nu}&=&\frac{1}{f_R}\bigg\{g_{\mu\nu}\left( \frac{f-R\
f_R}{2}\right)+(\nabla_\mu \nabla_\nu f_R-g_{\mu\nu}\Box f_R)-\rho
g_{\mu\nu}f_{\textbf{T}^{2}}T\\\label{7} &+&
(T+2\rho)f_{\textbf{T}^{2}}T_{\mu\nu}-2T^{\alpha}_\mu
T_{\alpha\nu}f_{\textbf{T}^{2}}\bigg\}.
\end{eqnarray}
The role of charge is determined through the electromagnetic
energy-momentum tensor given as
\begin{equation}\nonumber
{E}_{\mu\nu}=\frac{1}{4\pi}\left(\mathit{F}^{\alpha}_\mu
\mathit{F}_{\nu\alpha}-
\frac{1}{4}g_{\mu\nu}\mathit{F}^{\alpha\beta}\mathit{F}_{\alpha\beta}\right).
\end{equation}
The tensorial formulation of Maxwell field equations is given as
\begin{eqnarray*}
\mathit{F}_{[\mu\nu;\lambda]}=0,\quad\mathit{F}^{\mu\nu}_{;\nu}=\mu_o\emph{J}^{\mu}.
\end{eqnarray*}
where $\mu_o$ is the magnetic permeability. The electromagnetic
four-current vector is defined as
$\emph{J}_i=\frac{\sigma}{\sqrt{g_{00}}}\frac{\emph{d}x_i}{\emph{d}x^{4}}=\sigma\emph{v}_i$,
where $\sigma$ is the charge density.

To analyze the compact structure, we consider the static spherical
spacetime as
\begin{equation}\label{7}
\emph{ds}^{2}=\mathcal{F}^{2}(r)\emph{dt}^{2}-
\mathcal{G}^{2}(r)\emph{dr}^{2}-\emph{r}^{2}\emph{d}\theta^{2}-\emph{r}^{2}
\sin^{2}\theta\emph{d}\Phi^{2}.
\end{equation}
Consequently, the four-vector and four-velocity have the following
forms
\begin{eqnarray*}
\emph{s}^{\mu}=(0,\frac{1}{\mathcal{G}},0,0),~\emph{v}^{\mu}=(\frac{1}{\mathcal{F}},0,0,0),
\end{eqnarray*}
which imply that $\emph{v}_\mu\emph{v}^{\mu}=1,~
\emph{s}_\mu\emph{s}^{\mu}=-1,~\emph{v}_\mu\emph{s}^{\mu}=0$. The
Maxwell field equations for the considered metric turn out to be
\begin{equation*}
\phi^{''}+(\frac{2}{r}-\frac{\mathcal{F}^{'}}{\mathcal{F}}-\frac{\mathcal{G}^{'}}{\mathcal{G}})\phi^{'}=4
\pi\sigma\mathcal{F}\mathcal{G}^{2},
\end{equation*}
where prime denotes derivative with respect to $r$. The integration
of the above equation yields
\begin{equation*}
\phi^{'}=\frac{\mathcal{F}\mathcal{G}\emph{q}(r)}{r^{2}},
\end{equation*}
where the total charge within the sphere is given by
$\emph{q}(r)=4\pi\int^{r}_0\sigma\mathcal{G}r^{2}\emph{d}r$. Taking
covariant differentiation of Eq.(\ref{2}), we obtain
\begin{equation}\label{8}
\kappa^{2}\nabla^{\mu}T^{(D)}_{\mu\nu}=\nabla^{\mu}(\Theta_{\mu\nu}f_{\textbf{T}^{2}})
-\frac{1}{2}g_{\mu\nu}\nabla^{\mu}f,
\end{equation}
which implies that conservation of the energy-momentum tensor does
not hold leading to the existence of an unknown force that causes
non-geodesic movement of particles. The EMSG field equations
corresponding to the line element in Eq.(\ref{7}) are
\begin{eqnarray}\label{9}
\frac{1}{r^{2}}-\frac{1}{\mathcal{G}^{2}}\left(\frac{1}{r^{2}}-
\frac{2\mathcal{G}^{'}}{r\mathcal{G}}\right)&=&\frac{1}{f_R}\left(\rho+\varphi+
\varphi_{00}-2\pi\mathbb{E}^{2}\right),\\\nonumber
-\frac{1}{r^{2}}+\frac{1}{\mathcal{G}^{2}}\left(\frac{1}{r^{2}}+
\frac{2\mathcal{F}^{'}}{r\mathcal{F}}\right)
&=&\frac{1}{f_R}\left\{(\emph{p}_r-\varphi+\varphi_{11})+\left(\rho^{2}+2\rho\emph{p}_r
\right.\right.\\\nonumber
&-&\left.\left.2\rho\emph{p}_\bot+\emph{p}^{2}_r-2\emph{p}_r\emph{p}_\bot\right)f_{\textbf{T}^{2}}+2\pi\mathbb{E}^{2}\right\},\\\label{10}\\\nonumber
\frac{1}{\mathcal{G}^{2}}\left(\frac{\mathcal{F}^{''}}{\mathcal{F}}-
\frac{\mathcal{F}^{'}\mathcal{G}^{'}}{\mathcal{F}\mathcal{G}}\right)
-\frac{1}{r\mathcal{G}^{2}}\left(\frac{\mathcal{G}^{'}}{\mathcal{G}}-
\frac{\mathcal{F}^{}}{\mathcal{F}}\right)&=&\frac{1}{f_R}\left\{(\emph{p}_\bot-\varphi+
\varphi_{22})+\left(\rho^{2}-\rho\emph{p}_r\right.\right.\\\label{11}
&+&\left.\left.\rho\emph{p}_\bot-\emph{p}_r\emph{p}_\bot\right)f_{\textbf{T}^{2}}-2\pi\mathbb{E}^{2}\right\},
\end{eqnarray}
where
\begin{eqnarray}\nonumber
\varphi&=&\frac{f-Rf_R}{2},\\\nonumber
\varphi_{00}&=&-\frac{f^{''}_R}{\mathcal{G}^{2}}-\frac{1}{\mathcal{G}^{2}}\left(\frac{2}{r}-\frac{\mathcal{G}^{'}}{\mathcal{G}}\right),\\\nonumber
\varphi_{11}&=&-\frac{f^{'}_R\mathcal{F}^{'}}{\mathcal{F}\mathcal{G}^{2}}-\frac{2f^{'}_R}{r\mathcal{G}^{2}},\\\nonumber
\varphi_{22}&=&-\frac{1}{\mathcal{G}^{2}}\left\{f^{''}_R+\left(\frac{\mathcal{F}^{'}}{\mathcal{F}}-\frac{\mathcal{G}^{'}}{\mathcal{G}}+\frac{1}{r}\right)f^{'}_R\right\},\\\nonumber
\mathbb{E}&=& \frac{\emph{q}(r)}{4\pi r^{2}},
\end{eqnarray}
where $\mathbb{E}$ is the electric field intensity.

\section{Physical Characteristics of Matter Distribution}

The Riemann tensor measures the curvature of spacetime and is
represented through the Ricci tensor, Ricci scalar and Weyl tensor
($\mathcal{C}^\varepsilon_{\mu\nu\lambda}$) as
\begin{equation}\label{12}
R^{\varepsilon}_{\mu\nu\lambda}=\mathcal{C}^\varepsilon_{\mu\nu\lambda}+\frac{1}{2}R^{\varepsilon}_\nu
g_{\mu\lambda}-\frac{1}{2}R_{\mu\nu}\delta^{\varepsilon}_\lambda+
\frac{1}{2}R_{\mu\lambda}\delta^{\varepsilon}_\nu-\frac{1}{2}R^{\varepsilon}_\lambda
g_{\mu\nu}-\frac{1}{6} R\left(\delta^{\varepsilon}_\nu
g_{\mu\lambda}-g_{\mu\nu}\delta^{\varepsilon}_\lambda\right).
\end{equation}
The Weyl tensor is the traceless component of the Riemann tensor
which gauges the tidal constrain on a body. Utilizing the observer's
four-velocity, it can be decomposed into magnetic
($\mathcal{H}_{\mu\nu}$) and electric ($\mathcal{E}_{\mu\nu}$) parts
as
\begin{eqnarray}\nonumber
\mathcal{H}_{\mu\nu}=\frac{1}{2}\eta_{\mu\alpha\varepsilon\beta}\mathcal{C}^{\varepsilon\beta}_{\nu\sigma}\emph{v}^{\alpha}\emph{v}^{\sigma},~
\mathcal{E}_{\mu\nu}=\mathcal{C}_{\mu\lambda\nu\gamma}\emph{v}^{\lambda}\emph{v}^{\gamma}.
\end{eqnarray}
Here $\eta_{\alpha\beta\mu\nu}$ represents the Levi-Civita tensor,
$g_{\iota\beta\lambda\nu}=g_{\iota\lambda}g_{\beta\nu}-g_{\iota\nu}g_{\beta\lambda}$
and
$\mathcal{C}_{\alpha\beta\epsilon\sigma}=(g_{\alpha\beta\mu\nu}g_{\pi\lambda\kappa\varrho}-\eta_{\alpha\beta\mu\nu}\eta_{\pi\lambda\kappa\varrho})\emph{v}^{\mu}\emph{v}^{\kappa}\mathcal{E}^{\nu\varrho}$.
For a spherically symmetric system, the magnetic part of the Weyl
tensor varnishes whereas the electric component is written in terms
of unit four-vector and projection tensor as
\begin{equation}\label{13}
\mathcal{E}_{\mu\nu}=\mathcal{E}\left(\emph{s}_\mu\emph{s}_\nu+\frac{1}{3}\emph{h}_{\mu\nu}\right),
\end{equation}
where $ \mathcal{E}=-\frac{1}{2\mathcal{F}\mathcal{G}^{3}r^{2}}
\left(\mathcal{F}^{''}\mathcal{G}r^{2}
-\mathcal{F}^{'}\mathcal{G}^{'}r^{2}-\mathcal{F}\mathcal{G}^{3}+
\mathcal{F}\mathcal{G}^{'}r-\mathcal{F}^{'}\mathcal{G}r+\mathcal{F}\mathcal{G}\right)$
with $\mathcal{E}^{\mu}_\mu\\=0,~
\mathcal{E}_{(\epsilon\delta)}=\mathcal{E}_{\epsilon\delta}$ and
$\mathcal{E}_{\mu\lambda}\emph{v}^{\lambda}=0$.

Utilizing the definitions of Misner-Sharp \cite{41} and Tolman mass
\cite{42}, we develop an association between the Weyl tensor and
mass function to explore some characteristics of the spherical
framework. The mass obtained through Misner-Sharp as well as
Tolman's definitions has the same values at the boundary. However,
these definitions provide the same estimates of mass within the
interior in the scenario of isotropic and homogeneous fluid only.
The formulation developed by Misner and Sharp under the impact of
charge is given as
\begin{eqnarray}\nonumber
\frac{2}{r}\emph{m}&=&R^{2}_{232}=\left(1-\frac{1}{\mathcal{G}^{2}}\right)=\frac{1}{r}\int^{r}_0
T^{(D)0}_0 \tilde{r}^{2}\emph{d}\tilde{r}
=\frac{1}{r}\int^{r}_0\frac{1}{f_R}\big(\rho+\varphi+\varphi_{00}\\\label{14}
&-&2\pi\mathbb{E}^{2}\big) \tilde{r}^{2}\emph{d}\tilde{r}.
\end{eqnarray}
Using the field equations along with Eq.(\ref{13}), the mass
function is rewritten as
\begin{eqnarray}\nonumber
\emph{m}&=&r^{3}\left[\frac{1}{6
f_R}\Big\{(\rho-\emph{p}_r+\emph{p}_\bot)+(\varphi+\varphi_{00}-\varphi_{11}
-\varphi_{22}-6\pi\mathbb{E}^{2})\right.\\\label{15}&+&\left.(3\rho\emph{p}_\bot-3\rho\emph{p}_r
+\emph{p}_r\emph{p}_\bot-\emph{p}^{2}_r)f_{\textbf{T}^{2}}+\frac{\mathcal{E}}{3}\Big\}\right].
\end{eqnarray}
Equation (\ref{15}) along with (\ref{14}) leads to
\begin{eqnarray}\nonumber
\mathcal{E}&=&-\frac{1}{2 r^{3}}\int^{r}_0
\tilde{r}^{3}\bigg\{\left(\frac{1}{f_R}\right)^{'}(\rho+\varphi+\varphi_{00}-
2\pi\mathbb{E}^{2})+\left(\frac{1}{f_R}\right)\left(\rho+\varphi\right.\\\nonumber
&+&\left.\varphi_{00}-
2\pi\mathbb{E}^{2})^{'}\right)\bigg\}\emph{d}\tilde{r}+\frac{1}{2
f_R}\Big\{(\emph{p}_r-
\emph{p}_\bot)+(\varphi_{11}-\varphi_{22}-4\pi\mathbb{E}^{2})\\\label{16}
&+&(3\rho\emph{p}_r-3\rho\emph{p}_\bot
-\emph{p}_r\emph{p}_\bot+\emph{p}^{2}_r)f_{\textbf{T}^{2}}\Big\}.
\end{eqnarray}
The above expression describes how Weyl tensor and physical
properties of the fluid (anisotropic pressure, inhomogeneous energy
density and total charge) are interlinked. Inserting Eq.(\ref{16})
in (\ref{15}) yields
\begin{eqnarray}\nonumber
\emph{m}(r)&=&\frac{r^{3}}{6}T^{(D)0}_0-\frac{1}{6}\int^{r}_0
\tilde{r}^{3}\bigg\{\left(\frac{1}{f_R}\right)^{'}(\rho+\varphi+\varphi_{00}-
2\pi\mathbb{E}^{2})+\left(\frac{1}{f_R}\right)\big(\rho+\varphi\\\label{17}
&+&\varphi_{00}- 2\pi\mathbb{E}^{2}\big)^{'}d\tilde{r}\bigg\},
\end{eqnarray}
which shows the association of mass function with energy
inhomogeneity in $f(R,\textbf{T}^{2})$ gravity. The above result
coincides with GR for vanishing $\varphi$ and $\varphi_{00}$.

A self-gravitating body is in equilibrium when the inward force of
gravity is balanced by the outward pressure. The
Tolman-Opphenheimer-Volkoff (TOV) equation is the analog of
hydrostatic equilibrium equation in GR. Bekenstein \cite{43}
determined an extension of TOV equation in 1971 for charged compact
objects. The TOV equation for charged anisotropic fluid distribution
in $f(R,\textbf{T}^{2})$ gravity is determined by using Eq.(\ref{8})
as
\begin{eqnarray}\nonumber
\emph{p}^{'}_r&=&\frac{1}{1+(\emph{p}_r+2\rho-2\emph{p}_\bot)f_{\textbf{T}^{2}}}
\left[2f_{\textbf{T}^{2}}\bigg\{\frac{\mathcal{F}^{'}}{\mathcal{F}}\left(-\rho^{2}+2\rho\emph{p}_\bot-
2\rho\emph{p}_r-\emph{p}^{2}_r-4\emph{p}^{2}_\bot\right.\right.\\\nonumber
&+&\left.\left.2\emph{p}_r\emph{p}_\bot\right)-\left(\rho^{'}(\rho+2\emph{p}_r-2\emph{p}_\bot)+\emph{p}^{'}_\bot(-2\rho+7\emph{p}_\bot
-2\emph{p}_r)\right)-\frac{2}{r}\left(3\rho\emph{p}_r\right.\right.\\\nonumber&-&\left.\left.3\rho\emph{p}_\bot-\emph{p}_r\emph{p}_\bot
+\emph{p}^{2}_r+4\emph{p}^{2}_\bot\right)\bigg\}
-\frac{2}{r}\left(\frac{\emph{q}\emph{q}^{'}}{8\pi
r^{4}}\right)-\left\{\frac{\rho+\emph{p}_r}{r(r-2\emph{m})}\bigg[\emph{m}+\frac{r^{3}}{f_R}\right.\right.\\\nonumber
&\times&\left.\left.\left((\emph{p}_r-\varphi+\varphi_{11})+\left(\rho^{2}+2\rho\emph{p}_r-2\rho\emph{p}_\bot+\emph{p}^{2}_r-2\emph{p}_r\emph{p}_\bot\right)f_{\textbf{T}^{2}}\right)\bigg]+\frac{2}{r}\right.\right.\\\label{18}
&\times&\left.\left.
\left((\emph{p}_r-\emph{p}_\bot)-\frac{\emph{q}^{2}}{4\pi
r^{4}}\right)\right\}\right].
\end{eqnarray}
The mass inside the boundary $\Sigma$ (with radius $r_\Sigma$) of a
spherically symmetric distribution is calculated through Tolman's
formula as \cite{42}
\begin{equation}\label{19}
\emph{m}_{Tol}=\int^{r_\Sigma}_0\frac{\tilde{r}^{2}\mathcal{F}\mathcal{G}}{2}\left(T^{(D)0}_0
-T^{(D)1}_1-2T^{(D)2}_2\right)\emph{d}\tilde{r},
\end{equation}
which, for the current setup, is expressed as
\begin{eqnarray}\nonumber
\emph{m}_{Tol}&=&\mathcal{F}\mathcal{G}\bigg[\emph{m}(r)+\frac{r^{3}}{2
f_R}\left\{(\emph{p}_r-\varphi+\varphi_{11})+\left(\rho^{2}
+2\rho\emph{p}_r-2\rho\emph{p}_\bot\right.\right.\\\label{20}
&+&\left.\left.\emph{p}^{2}_r-2\emph{p}_r\emph{p}_\bot
\right)f_{\textbf{T}^{2}}+2\pi\mathbb{E}^{2}\right\}\bigg].
\end{eqnarray}
Using the field equations, the above expression reduces to
\begin{equation}\nonumber
\emph{m}_{Tol}=\frac{2\mathcal{F}^{'}}{\mathcal{G}}\left(\frac{r^{2}}{2}\right).
\end{equation}
The gravitational acceleration of a test particle in a static
gravitational field is related to Tolman mass as
\begin{equation*}
\emph{a}=\frac{\mathcal{F}^{'}}{\mathcal{F}\mathcal{G}}=\frac{\emph{m}_{Tol}}{\mathcal{F}r^{2}}.
\end{equation*}
The above expression describes the interpretation of
$\emph{m}_{Tol}$ as active gravitational mass. The expression for
Tolman mass can be rewritten using Eq.(\ref{16}) as
\begin{eqnarray}\nonumber
\emph{m}_{Tol}&=&(\emph{m}_{Tol})_\Sigma\left(\frac{r}{r_\Sigma}\right)^{3}+r^{3}\int^{r_\Sigma}_r
\frac{\mathcal{F}\mathcal{G}}{\tilde{r}}\left[\frac{1}{2f_R}\left\{(\emph{p}_r-
\emph{p}_\bot)+(\varphi_{11}-\varphi_{22})\right.\right.\\\label{21}
&+&\left.\left.4\pi\mathbb{E}^{2}+(3\rho\emph{p}_r-3\rho\emph{p}_\bot+\emph{p}^{2}_r-\emph{p}_r\emph{p}_\bot)
f_{\textbf{T}^{2}}\right\}+\mathcal{E}\right]\emph{d}\tilde{r}.
\end{eqnarray}
The integral term shows that the Tolman mass depends mainly on
anisotropy, inhomogeneous energy density, electromagnetic field and
non-linear combination of $f(R,\textbf{T}^{2})$ dark source terms.

\section{Structure Scalars}

Herrera et al. \cite{44} formulated a methodology for orthogonal
decomposition of the Riemann tensor to obtain structure scalars.
Adopting his technique, we consider the following tensor quantities
\begin{eqnarray}\label{22}
\mathcal{Y}_{\mu\nu}&=&R_{\mu\gamma\nu\delta}\emph{v}^{\gamma}\emph{v}^{\delta},\\\label{23}
\mathcal{Z}_{\mu\nu}&=&
^{*}R_{\mu\gamma\nu\delta}\emph{v}^{\gamma}\emph{v}^{\delta}=
\frac{1}{2}\eta_{\mu\gamma\lambda\varsigma}R^{\lambda\varsigma}_{\nu\delta}\emph{v}^{\gamma}\emph{v}^{\delta},\\\label{24}
\mathcal{X}_{\mu\nu}&=&^{*}R^{*}_{\mu\gamma\nu\delta}\emph{v}^{\gamma}\emph{v}^{\delta}
=\frac{1}{2}\eta^{\lambda\varsigma}_{\mu\gamma}R^{*}_{\lambda\varsigma\nu\delta}\emph{v}^{\gamma}\emph{v}^{\delta},
\end{eqnarray}
where $*$ denotes the dual tensor defined as
$R^{*}_{\mu\nu\emph{e}\delta}=\frac{1}{2}\eta_{\lambda\omega\emph{e}\delta}R^{\lambda\omega}_{\mu\nu}$.
Using Eq.(\ref{12}), we can write the Riemann tensor as
\begin{equation}\label{25}
R^{\mu\gamma}_{\nu\delta}=\mathcal{C}^{\mu\gamma}_{\nu\delta}+2T^{(D)[\mu}_{[\nu}\delta^{\gamma]}_{\delta]}
+T^{(D)}\left(\frac{1}{3}\delta^{\mu}_{[\nu}\delta^{\gamma}_{\delta]}-
\delta^{[\mu}_{[\nu}\delta^{\gamma]}_{\delta]}\right).
\end{equation}
The Riemann tensor can be split using the above expression as
\begin{equation}\nonumber
R^{\mu\gamma}_{\nu\delta}=R^{\mu\gamma}_{\mathbb{(I)}\nu\delta}+R^{\mu\gamma}_{\mathbb{(II)}\nu\delta}
+R^{\mu\gamma}_{\mathbb{(III)}\nu\delta},
\end{equation}
where
\begin{eqnarray}\nonumber
R^{\mu\gamma}_{\mathbb{(I)}\nu\delta}&=&\frac{2}{f_R}\left[\nabla^{[\mu}\nabla_{[\nu}\delta^{\gamma]}_{\delta]}+
\left\{(\rho^{2}-\rho\emph{p}_r+\rho\emph{p}_\bot-\emph{p}_r\emph{p}_\bot)f_{\textbf{T}^{2}}+(\rho+\emph{p}_\bot)\right\}\right.\\\nonumber
&\times&\left.\emph{v}^{[\mu}\emph{v}_{[\nu}\delta^{\gamma]}_{\delta]}+
\left\{(-\rho^{2}+\rho\emph{p}_r-\rho\emph{p}_\bot+\emph{p}_r\emph{p}_\bot)f_{\textbf{T}^{2}}+(\varphi-\emph{p}_\bot-\Box
f_R)\right\}\right.\\\nonumber
&\times&\left.\delta^{[\mu}_{[\nu}\delta^{\gamma]}_{\delta]}+\left\{\left(\rho^{2}+3\rho\emph{p}_r-3\rho\emph{p}_\bot-
\emph{p}_r\emph{p}_\bot\right)f_{\textbf{T}^{2}}+(\emph{p}_r-\emph{p}_\bot)\right\}\emph{s}^{[\mu}\emph{s}_{[\nu}\delta^{\gamma]}_{\delta]}\right.\\\nonumber
&+&\left.
\emph{E}^{[\mu}_{[\nu}\delta^{\gamma]}_{\delta]}\right],\\\nonumber
R^{\mu\gamma}_{\mathbb{(II)}\nu\delta}&=&\frac{2}{3 f_R}\left\{3\Box
f_R-4\varphi-(\rho-\emph{p}_r-2\emph{p}_\bot)+(3\rho^{2}-4\emph{p}_r\emph{p}_\bot-\emph{p}^{2}_r)f_{\textbf{T}^{2}}\right\}\\\nonumber
&\times&
\left(\delta^{[\mu}_{[\nu}\delta^{\gamma]}_{\delta]}\right),\\\nonumber
R^{\mu\gamma}_{\mathbb{(III)}\nu\delta}&=&4\emph{v}^{[\mu}\emph{v}_{[\nu}\mathcal{E}^{\gamma]}_{\delta]}
-\epsilon^{\mu\gamma}_\alpha\epsilon_{\nu\delta\beta}\mathcal{E}^{\alpha\beta},
\end{eqnarray}
where
$\eta_{\mu\alpha\nu\beta}=\emph{v}_\mu\epsilon_{\alpha\nu\beta},~
\epsilon_{\mu\nu\alpha}\emph{v}^{\alpha}=0$. The structure scalars
are very helpful in exploring the physical properties of a system as
they are a combination of physical parameters that are useful in
determining the complexity of a system.

We can formulate $\mathcal{X}_{\mu\nu}$, $\mathcal{Y}_{\mu\nu}$ and
$\mathcal{Z}_{\mu\nu}$ in terms of physical variables using the
Riemann tensor. The quantities $\mathcal{X}_{\mu\nu}$ and
$\mathcal{Y}_{\mu\nu}$ can be expressed in terms of their trace-free
$(\mathcal{X}_{TF},~\mathcal{Y}_{TF})$ and trace parts
$(\mathcal{X}_{T}=\mathcal{X}^{\mu}_\mu,~\mathcal{Y}_{T}=\mathcal{Y}^{\mu}_\mu)$
as
\begin{eqnarray}\nonumber
\mathcal{X}_{\mu\nu}&=&\frac{\mathcal{X}_T\emph{h}_{\mu\nu}}{3}+
\mathcal{X}_{TF}\left(\emph{s}_\mu\emph{s}_\nu+\frac{1}{3}\emph{h}_{\mu\nu}\right),\\\nonumber
\mathcal{Y}_{\mu\nu}&=&\frac{\mathcal{Y}_T\emph{h}_{\mu\nu}}{3}+
\mathcal{Y}_{TF}\left(\emph{s}_\mu\emph{s}_\nu+\frac{1}{3}\emph{h}_{\mu\nu}\right).
\end{eqnarray}
The trace and trace-free parts in the context of
$f(R,\textbf{T}^{2})$ gravity are calculated as
\begin{eqnarray}\label{26}
\mathcal{X}_T&=&\frac{3\Box f_R}{2
f_R}-\frac{\varphi}{f_R}-\frac{\emph{q}^{2}}{8\pi
r^{4}f_R}-\frac{1}{f_R}(\rho+\frac{11}{2}\emph{p}_r\emph{p}_\bot)f_{\textbf{T}^{2}},\\\nonumber
\mathcal{Y}_T&=&-\frac{\varphi}{f_R}+\frac{1}{2f_R}\Big[\rho+3\emph{p}+3\Box
f_R+\frac{\emph{q}^{2}}{4\pi
r^{4}}+\left(3\rho^{2}+\emph{p}^{2}_r-6\emph{p}_r\emph{p}_\bot\right)f_{\textbf{T}^{2}}\Big],\\\label{27}\\\nonumber
\mathcal{X}_{TF}&=&-\mathcal{E}-\frac{1}{2f_R}\Big[\left(\emph{p}_\bot-\emph{p}_r\right)+\frac{\emph{q}^{2}}{4\pi
r^{4}}+\left(3\rho\emph{p}_\bot-3\rho\emph{p}_r-\emph{p}^{2}_r+\emph{p}_r\emph{p}_\bot\right)f_{\textbf{T}^{2}}\Big],\\\label{28}\\\nonumber
\mathcal{Y}_{TF}&=&\mathcal{E}+\frac{1}{2f_R}\Big[\left(\emph{p}_r-\emph{p}_\bot\right)-\frac{\emph{q}^{2}}{4\pi
r^{4}}+\left(\emph{p}^{2}_r+3\rho\emph{p}_r-3\rho\emph{p}_\bot-\emph{p}_r\emph{p}_\bot\right)f_{\textbf{T}^{2}}\Big].\\\label{29}
\end{eqnarray}
It is noted from Eq.(\ref{28}) that the scalar $\mathcal{X}_{TF}$ is
associated with the impact of total charge and inhomogeneous energy
density of matter distribution. The scalar $\mathcal{Y}_T$ analyzes
the impact of principal stresses generated by density inhomogeneity
while the total energy density of the structure is determined
through $\mathcal{X}_T$ in the presence of charge. The physical
importance of scalar $\mathcal{Y}_{TF}$ can be interpreted by
utilizing Eqs.(\ref{21}) and (\ref{29}) as
\begin{equation}\label{30}
\emph{m}_{Tol}=(\emph{m}_{Tol})_\Sigma\left(\frac{r}{r_\Sigma}\right)^{3}-r^{3}\int^{r_\Sigma}_r
\frac{\mathcal{F}\mathcal{G}}{\tilde{r}}\left[\frac{1}{2f_R}(\varphi_{22}-\varphi_{11})
-\mathcal{Y}_{TF}\right]\emph{d}\tilde{r}.
\end{equation}
Equations (\ref{29}) and (\ref{30}) show that the scalar
$\mathcal{Y}_{TF}$ determines the effect of inhomogeneous energy
density, anisotropic pressure and charge on the Tolman mass in the
presence of correction terms in EMSG.  We obtain local anisotropic
pressure and electric charge with dark source terms of
$f(R,\textbf{T}^{2})$ theory as
\begin{eqnarray}\nonumber
\mathcal{Y}_{TF}+\mathcal{X}_{TF}&=&\frac{1}{f_R}\left[(\emph{p}_r-\emph{p}_\bot)+
(\emph{p}^{2}_r+3\rho\emph{p}_r-3\rho\emph{p}_\bot-\emph{p}_r\emph{p}_\bot)f_{\textbf{T}^{2}}-\frac{\emph{q}^{2}}{4\pi
r^{4}}\right].
\end{eqnarray}

\section{Complexity Factor}

Many factors are responsible for creating complexity in a stellar
structure. Such factors include heat dissipation, electromagnetic
field, inhomogeneity, pressure anisotropy, viscosity, etc. In
general, any structure possessing homogeneous energy together with
isotropic pressure is considered as the only framework with
insignificant complexity. In the considered setup, complexity is
caused by energy density inhomogeneity, pressure anisotropy,
electromagnetic field and correction terms of $f(R,\textbf{T}^{2})$
gravity. The structure scalar $\mathcal{Y}_{TF}$ connects the
sources of complexity and also measures their impact on Tolman mass.
Thus, $\mathcal{Y}_{TF}$ is a suitable candidate for the complexity
factor of the considered system. Substituting Eq.(\ref{16}) in
(\ref{29}) yields $\mathcal{Y}_{TF}$ in terms of state variables as
\begin{eqnarray}\nonumber
\mathcal{Y}_{TF}&=&\frac{1}{2f_R}\bigg\{2(\emph{p}_r-\emph{p}_\bot)+\varphi_{11}-\varphi_{22}-\frac{\emph{q}^{2}}{2\pi
r^{4}}+\left(2\emph{p}^{2}_r+6\rho\emph{p}_r-6\rho\emph{p}_\bot\right.\\\nonumber
&-&\left.2\emph{p}_r\emph{p}_\bot\right)f_{\textbf{T}^{2}}\bigg\}-\frac{1}
{2r^{3}}\int^{r}_0
\tilde{r}^{3}\left\{\left(\frac{1}{f_R}\right)^{'}\left(\rho+\varphi+\varphi_{00}-
\frac{\emph{q}^{2}}{8\pi \tilde{r}^{4}}\right)\right.\\\label{31}
&+&\left.\left(\frac{1}{f_R}\right)\left(\rho+\varphi+\varphi_{00}-
\frac{\emph{q}^{2}}{8\pi
\tilde{r}^{4}}\right)^{'}d\tilde{r}\right\}.
\end{eqnarray}
We proceed by assuming the following expression for charge \cite{45}
\begin{equation}\label{32}
\emph{q}(r)=\emph{q}_o(r)\left(\frac{r}{r_\Sigma}\right)^{3}=\beta
r^{3},
\end{equation}
where subscript $o$ denotes the value of the physical quantity at
$r=0$ and $\beta=\frac{\emph{q}_o(r)}{r_\Sigma^{3}}$. From
Eqs.(\ref{31}) and (\ref{32}), we deduce that complexity decreases
in the presence of charge.

In $f(R,\textbf{T}^{2})$ theory, five unknowns
$\left(\mathcal{F},\mathcal{G},\rho,\emph{p}_r,\emph{p}_\bot\right)$
are present in the system of field equations. We therefore require
additional conditions to obtain a solution. For this purpose, one
constraint is obtained through the vanishing complexity factor. By
setting Eq.(\ref{31}) equal to zero, we acquire the vanishing
complexity condition as
\begin{eqnarray}\nonumber
\Pi&=&\frac{1}{1+(3\rho+\emph{p}_r)f_{\textbf{T}^{2}}}\left[\frac{1}
{2r^{3}}\int^{r}_0
\tilde{r}^{3}\left\{\left(\frac{1}{f_R}\right)^{'}\left(\rho+\varphi+\varphi_{00}-
\frac{\emph{q}^{2}}{8\pi
\tilde{r}^{4}}\right)\right.\right.\\\label{33}
&+&\left.\left.\left(\frac{1}{f_R}\right)\left(\rho+\varphi+\varphi_{00}-
\frac{\emph{q}^{2}}{8\pi
\tilde{r}^{4}}\right)^{'}\right\}\emph{d}\tilde{r}+\frac{\left(\varphi_{22}-\varphi_{11}\right)}{2}+\frac{\emph{q}^{2}}{4\pi
r^{4}}\right].
\end{eqnarray}
The complexity factor vanishes for isotropic and homogeneous fluid
distribution in GR. However, in $f(R,\textbf{T}^{2})$, the
complexity of a stellar system with homogeneous and isotropic matter
configuration vanishes if the system obeys the following condition
\begin{eqnarray}\nonumber
&&\int^{r}_0
\tilde{r}^{3}\bigg\{\left(\frac{1}{f_R}\right)^{'}\left(\rho+\varphi+\varphi_{00}-
\frac{\emph{q}^{2}}{8\pi \tilde{r}^{4}}\right)
+\left(\frac{1}{f_R}\right)\Big(\varphi+\varphi_{00}\\\label{34} &&-
\frac{\emph{q}^{2}}{8\pi
\tilde{r}^{4}}\Big)^{'}\bigg\}\emph{d}\tilde{r}
+\frac{1}{2f_R}(\varphi_{22}-\varphi_{11})+\frac{1}{f_R}\frac{\emph{q}^{2}}{4\pi
r^{4}}=0.
\end{eqnarray}
We now evaluate the vanishing complexity condition for a specific
EMSG model given as \cite{34}
\begin{equation}\label{35}
f\left(R,\textbf{T}^{2}\right)=R+\zeta \textbf{T}^{2}.
\end{equation}
For the above model, the vanishing complexity condition reduces to
\begin{eqnarray}\nonumber
\Pi&=&\frac{1}{1+(3\rho+\emph{p}_r)\zeta}\left[\frac{1}{2r^{3}}\int^{r}_0\tilde{r}^{3}\rho^{'}\emph{d}\tilde{r}
+\left(1-\frac{1}{4\pi}\right)\frac{\emph{q}^{2}}
{r^{4}}\right.\\\label{36}&+&\left.\frac{\zeta}{4r^{3}}\int^{r}_0\tilde{r}^{3}\left(\rho^{2}+\emph{p}^{2}_r+
2\emph{p}^{2}_\bot\right)^{'}\emph{d}\tilde{r}\right].
\end{eqnarray}
Even after employing the condition $\mathcal{Y}_{TF}=0$, we still
require a condition to solve the field equations. For this purpose,
we utilize the energy density of Gokhroo-Mehra solution as well as
polytropic equation of state to obtain the corresponding solutions.

\subsection{The Gokhroo-Mehra Solution}

Gokhroo and Mehra \cite{46} considered a specific form of energy
density to compute the solutions of the field equations representing
an anisotropic spherical structure. They formulated a model that
explained greater red-shifts of various quasi-stellar system as well
as the dynamics of neutron stars. For the considered system, we will
assume the form of energy density proposed in \cite{46} and
determine the behavior of compact structures by incorporating the
condition of disappearing complexity in the presence of charge. The
assumed energy density is
\begin{equation}\label{37}
\rho=\rho_o\left(1-\frac{\emph{H}r^{2}}{r^{2}_\Sigma}\right),
\end{equation}
where $\rho_o$ is constant and $\emph{H}\in (0,1)$. Employing
Eqs.(\ref{14}), (\ref{35}) and (\ref{37}), we have
\begin{eqnarray}\nonumber
\emph{m}(r)&=&\frac{\rho_or^{3}}{6}-\frac{\emph{H}\rho_o
r^{5}}{10r^{2}_\Sigma}+\frac{\zeta\rho^{2}_or^{3}}{6}\left(\frac{1}{2}
-\frac{3\emph{H}r^{2}}{5r^{2}_\Sigma}+\frac{3\emph{H}^{2}r^{4}}{14r^{4}_\Sigma}\right)\\\label{38}
&+&\frac{\zeta}{4}\int^{r}_0
\tilde{r}^{2}(\emph{p}^{2}_r+2\emph{p}^{2}_\bot)\emph{d}\tilde{r}-\frac{\emph{q}^{2}}{16\pi
r}+\frac{1}{8\pi}\int^{r}_0\frac{\emph{q}\emph{q}^{'}}{\tilde{r}}\emph{d}\tilde{r},
\end{eqnarray}
which leads to
\begin{eqnarray}\nonumber
\mathcal{G}^{-2}&=&1-\frac{\rho_or^{2}}{3}+\frac{\emph{H}\rho_or^{4}}{5r^{2}_\Sigma}-
\zeta\left(\frac{\rho^{2}_or^{2}}{6}+\frac{\emph{H}^{2}\rho^{2}_o
r^{6}}{14r^{4}_\sigma} -\frac{\emph{H}\rho^{2}_o
r^{4}}{5r^{2}_\sigma}\right)\\\nonumber
&-&\frac{\zeta}{2r}\int^{r}_0\tilde{r}^{2}(\emph{p}^{2}_r+2\emph{p}^{2}_\bot)\emph{d}\tilde{r}
+\frac{\emph{q}^{2}}{8\pi r^{2}}-\frac{1}{4\pi
r}\int^{r}_0\frac{\emph{q}\emph{q}^{'}}{\tilde{r}}\emph{d}\tilde{r}.
\end{eqnarray}

We introduce new variables (to determine the unknowns) as
\begin{eqnarray}\nonumber
\mathcal{F}^{2}(r)=\emph{e}^{\int\left(2\emph{z}(r)-\frac{2}{r}\right)\emph{d}r},\quad
\frac{1}{\mathcal{G}^{2}}=\emph{x}(r).
\end{eqnarray}
From Eqs.(\ref{10}) and (\ref{11}), we obtain
\begin{eqnarray}\nonumber
\Pi\left[1+(3\rho+\emph{p}_r)\zeta\right]&=&\frac{1}{\mathcal{G}^{2}}\left[\frac{1}{r^{2}}-\frac{\mathcal{F}^{''}}{\mathcal{F}}+\frac
{\mathcal{F}^{'}}{\mathcal{F}r}+\frac{\mathcal{G}^{'}}{\mathcal{G}}\left(\frac{\mathcal{F}}{\mathcal{F}}+\frac{1}{r}\right)\right]-\frac{1}{r^{2}}-\frac{\emph{q}^{2}}{4\pi
r^{4}}.
\end{eqnarray}
After inserting new variables, the above equation is rewritten as
\begin{equation}\nonumber
\emph{x}^{'}+\emph{x}\left[\frac{2\emph{z}^{'}}{\emph{z}}+2\emph{z}-
\frac{6}{r}+\frac{4}{r^{2\emph{z}}}\right]+\frac{\emph{q}^{2}}{2\pi
r^{4}\emph{z}}=-\frac{2}{\emph{z}}\left[\frac{1}{r^{2}}+\Pi\left\{1+\zeta(3\rho+\emph{p}_r)\right\}\right],
\end{equation}
whose integration yields the radial metric function as
\begin{equation}\nonumber
\mathcal{G}^{2}(r)=\frac{\emph{e}^{2\int\left(\frac{2}{{r}^{2}\emph{z}({r})}
+\emph{z}(r)\right)\emph{d}r}\emph{z}^{2}(r)}{r^{6}\left[-2\int\frac{\emph{e}^{2\int\left(\frac{2}{{r}^{2}\emph{z}(r)}
+\emph{z}(r)\right)\emph{d}{r}}\emph{z}(r)(1+\Pi(r)[1+\zeta(3\rho+\emph{p}_r)])}{r^{8}}\emph{d}r-\int\frac{\emph{q}^{2}}{2\pi
r^{4}\emph{z}(r)}\emph{d}(r)+\mathbb{C}\right]},
\end{equation}
where $\mathbb{C}$ is the constant of integration. Thus, the line
element can be written in terms of $\emph{z}(r)$ and $\Pi$ as
\begin{eqnarray}\nonumber
\emph{ds}^{2}&=&
\frac{-\emph{e}^{2\int\left(\frac{2}{r^{2}\emph{z}(r)}
+\emph{z}(r)\right)\emph{d}r}\emph{z}^{2}(r)\emph{dr}^{2}}{r^{6}\left[-2\int\frac{\emph{e}^{2\int\left(\frac{2}{r^{2}\emph{z}(r)}
+\emph{z}(r)\right)\emph{d}r}\emph{z}(r)(1+\Pi(r)[1+\zeta(3\rho+\emph{p}_r)])}{r^{8}}\emph{d}r-\int\frac{\emph{q}^{2}}{2\pi
r^{4}\emph{z}(r)}\emph{d}(r)+\mathbb{C}\right]}\\\label{38}
&-&r^{2}\emph{d}\theta^{2}-r^{2}{\sin}^{2}\theta\emph{d}\Phi^{2}+\emph{e}^{\int\left(2\emph{z}(r)-\frac{2}{r}\right)\emph{d}r}\emph{dt}^{2}.
\end{eqnarray}

\subsection{Polytropic Equation of State}

Various physical variables have different roles in determining the
interior of self-gravitating structures. However, some variables
play a more dominant role in analyzing the structure than others. An
equation of state that effectively determines the combination of
vital variables assists in the analysis of stellar structures. The
polytropic equation of state, defining the relation of energy
density with radial pressure, has widely been used to study
anisotropic stellar objects \cite{47*}. The polytropic equation of
state for anisotropic fluid distribution is
\begin{equation}\nonumber
\emph{p}_r=\mathcal{K}\rho^{\gamma}=\mathcal{K}\rho^{1+\frac{1}{n}},
\end{equation}
where $\gamma$ is the polytropic exponent, the polytropic constant
is represented by $\mathcal{K}$ and $n$ denotes the polytropic
index. We introduce the following variables to determine the
dimensionless forms of TOV equation and mass function
\begin{eqnarray}\nonumber
\omega=\frac{\emph{p}_{ro}}{\rho_o},\quad
r=\frac{\xi}{\mathcal{A}},\quad
\mathcal{A}^{2}=\frac{\rho_o}{2\omega(n+1)},\quad\Omega(\xi)^{n}=\frac{\rho}{\rho_o},\quad\Upsilon(\xi)
=\frac{2\emph{m}(r)\mathcal{A}^{3}}{\rho_o},
\end{eqnarray}
where $\omega,~\xi,~\Omega$ and $\Upsilon$ are dimensionless
variables. Substituting these variables in TOV equation and mass
function, we obtain their respective dimensionless forms as
\begin{eqnarray}\nonumber
&&\frac{\left(1-\frac{2(n+1)\omega\Upsilon}{\xi}+
\frac{\emph{q}^{2}\rho_o}{4\pi\omega(n+1)\xi^{2}}\right)}{(1+\omega\Omega)}
\left[\left\{(-\rho_o\Omega^{n}-2\Pi-7\omega^{2}\Omega^{n+2}-\frac{8\Pi^{2}}
{\rho_o\Omega^{n}})\right\}^{-1}\right.\\\nonumber
&&\left.\left\{\left(\xi^{2}\frac{\mathit{d}\Omega}{\mathit{d}\xi}\right)
\left(1+(2\Pi-\omega\rho_o\Omega^{n+1}+2\rho_o\Omega^{n})
2\zeta\right)-\frac{2\zeta\xi^{2}}{\omega(n+1)}\left[-\rho_on\Omega^{n-1}\frac
{\mathit{d}\Omega}{\mathit{d}\xi}\right.\right.\right.\\\nonumber
&&\left.\left.\left.+2\omega\rho_o(n+1)\Omega^{n}\frac{\mathit{d}\Omega}
{\mathit{d}\xi}
-2n\Omega^{-1}\Pi\frac{\mathit{d}\Omega}{\mathit{d}\xi}-2\frac{\mathit{d}\Pi}{\mathit{d}\xi}
-12\omega^{2}\rho_o(n+1)\Omega^{n+1}\frac{\mathit{d}\Omega}{\mathit{d}\xi}\right.\right.\right.\\\nonumber
&&\left.\left.\left.+12\omega\Omega\frac{\mathit{d}\Pi}{\mathit{d}\xi}+14\omega(n+1)\Pi\frac{\mathit{d}\Omega}
{\mathit{d}\xi}
-\frac{14\Pi}{\rho_o\Omega^{n}}\frac{\mathit{d}\Pi}{\mathit{d}\xi}
+\frac{2}{\xi}\left(3\Pi-15\omega\Omega\Pi-\frac{8\Pi^{2}}{\rho_o\Omega^{n}}\right.\right.\right.\right.\\\nonumber
&&\left.\left.\left.\left.-8\omega^{2}\rho_o\Omega^{n+2}\right)
\right]+\frac{2\Pi\xi\Omega^{-n}}{\omega\rho_o(n+1)}-
\frac{\omega^{2}\rho_o\emph{q}\frac{\emph{dq}}{\emph{d}\xi}\Omega^{-n}}{4\pi\xi^{2}(n+1)^{2}}\right\}\right]
-\frac{\xi^{3}\zeta}{\rho_o}\left[\rho^{2}_o\Omega^{n}\left(\frac{1}{2}\Omega^{n}\right.\right.\\\nonumber
&&\left.\left.+\frac{1}{2}\omega^{2}\Omega-\omega^{2}\Omega^{n+2}-\omega^{2}\Omega^{n+1}\right)+4\omega
\rho_o\Pi\Omega^{n+1}\right]+\Upsilon+\omega\xi^{3}\Omega^{n+1}\Big\{1+\frac{\zeta}{2}\\\label{39}
&&\times
\Big(\rho_o\Omega^{n}+3\omega^{2}\rho_o\Omega^{n+2}+\frac{2\Pi^{2}}{\rho_o\Omega^{n}}-4\omega\Omega\Pi
\Big)\Big\}-\frac{\emph{q}^{2}\rho_o}{4\pi\xi\omega^{2}(n+1)^{2}}=0,\\\nonumber
&&\frac{\mathit{d}\Upsilon}{\mathit{d}\xi}=\xi^{2}\Omega^{n}\Bigg\{1-\frac{\emph{q}^{2}\rho_o}{4\pi\Omega^{n}(n+1)^{2}\xi^{4}}+\frac{\zeta}{2}
\Big(\rho_o\Omega^{n}+3\omega^{2}\rho_o\Omega^{n+2}+\frac{2\Pi^{2}}{\rho_o\Omega^{n}}\\\label{40}
&&-4\omega\Omega\Pi\Big)\Bigg\}.
\end{eqnarray}
Equations (\ref{39}) and (\ref{40}) constitute a system of
differential equations consisting of three unknown functions
$\Omega(\xi),\Pi(\xi)$ and $\Upsilon(\xi)$. To evaluate a unique
solution of this system, we impose the condition of vanishing
complexity. The vanishing complexity condition is written in
dimensionless variables as
\begin{eqnarray}\nonumber
&&\frac{6\pi}{n\rho_o}\left(1+(3+\omega\Omega)\rho_o\Omega^{n}\zeta\right)+\frac{2\xi}{n\rho_o}
\left[1+(3+\omega\Omega)\rho_o\zeta\Omega^{n}-\omega\zeta\rho_o\Omega^{n+1}\right.\\\nonumber
&&\left.+\zeta\Pi\right]\frac{\mathit{d}\Pi}{\mathit{d}\xi}=\bigg[\frac{-2\xi\Pi}{n\rho_o}
\left(3n\rho_o\Omega^{n-1}+\omega\rho_o(n+1)\Omega^{n}\right)\zeta+\xi\Omega^{n-1}+\frac{\zeta\xi}{n}\\\nonumber
&&\times\left(n\rho_o\Omega^{2n-1}+3\omega^{2}\rho_o(1+n)\Omega^{2n+1}-2\omega(n+1)\Omega^{n}\Pi\right)\bigg]
\frac{\mathit{d}\Omega}{\mathit{d}\xi}-\frac{1}{4\pi
n\rho_o}\\\label{41}&&\times\left(\frac{\emph{q}\emph{q}^{'}}{\xi^{4}}-\frac{4\emph{q}^{2}}{\xi^{5}}\right).
\end{eqnarray}

We obtain a unique solution for a complexity-free spherical stellar
structure for some specific values of the parameter $n$ and
$\omega$. The numerical solution is obtained by solving
Eqs.(\ref{39})-(\ref{41}) together with the initial conditions
$\Upsilon(0)=0,~\Omega(0)=1,~ \Pi(0)=0$ \cite{47}. It is essential
for a physically valid model that the state parameters (such as
energy density, pressure) should be finite, maximum and positive at
its center. Moreover, they should follow a monotonically decreasing
behavior towards the boundary. Also, the mass function should be a
positive and increasing function of the radial coordinate.
\begin{figure}\center
\epsfig{file=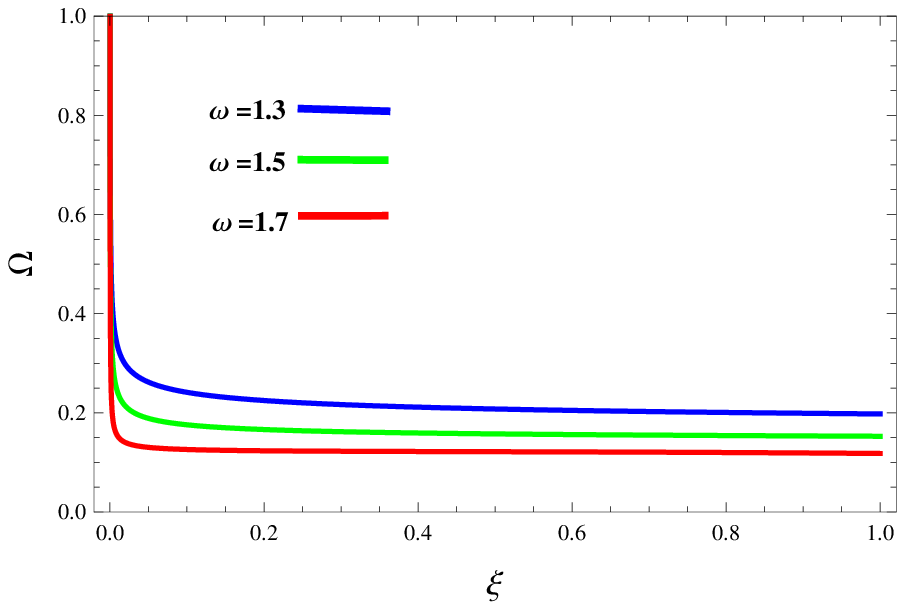,width=0.5\linewidth}\epsfig{file=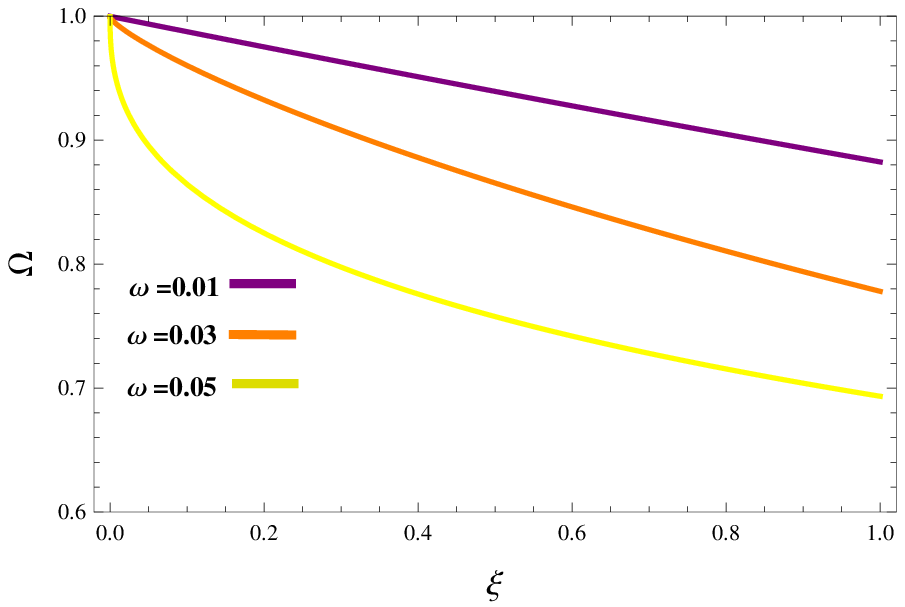,width=0.5\linewidth}
\caption{Plots of energy density versus $\xi$ for $n=3$, $\rho_o=5$,
$\zeta=7$, $\beta=0.001$.}
\end{figure}
\begin{figure}
\epsfig{file=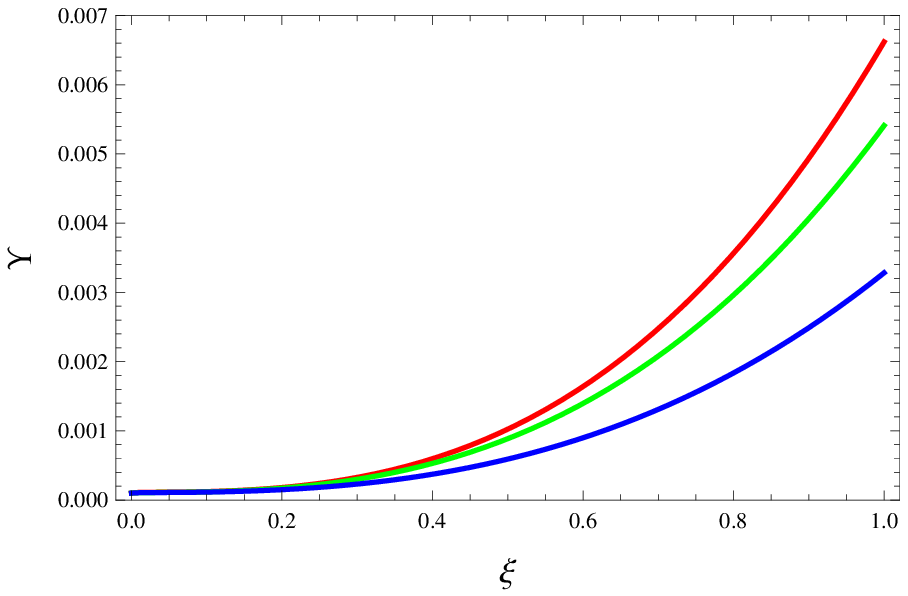,width=0.5\linewidth}\epsfig{file=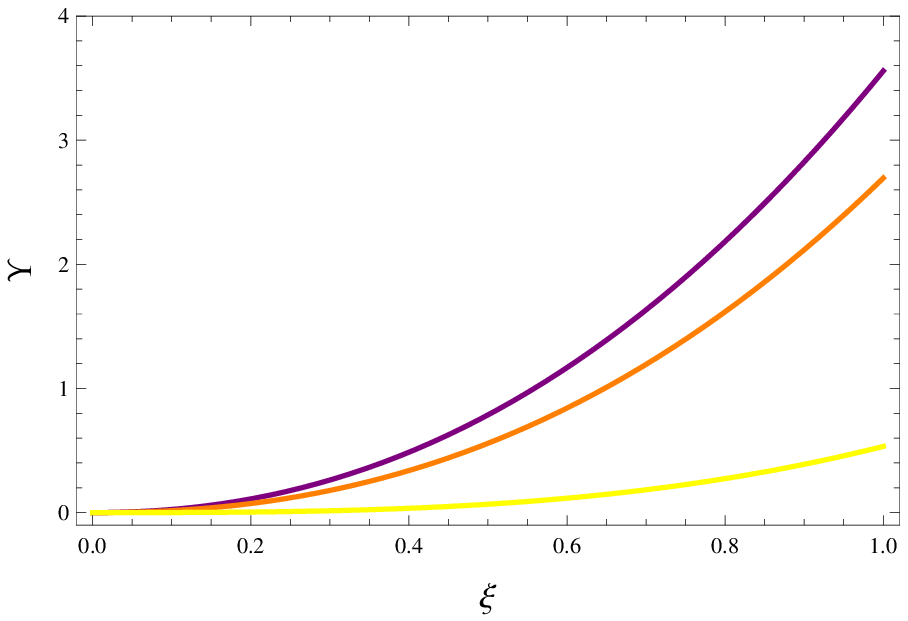,width=0.48\linewidth}
\caption{Plots of mass function versus $\xi$ for $n=3$, $\rho_o=5$,
$\zeta=7$, $\beta=0.001$.}
\end{figure}
\begin{figure}
\epsfig{file=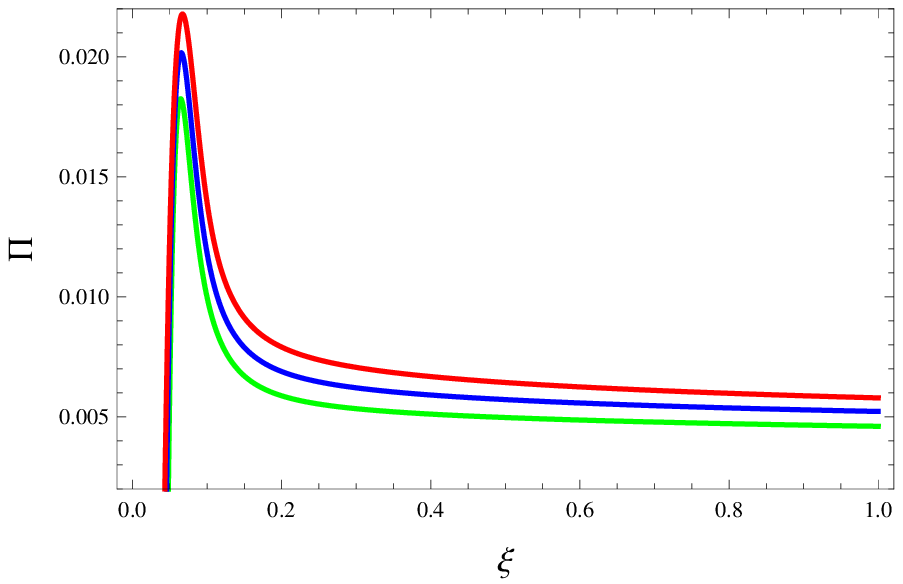,width=0.5\linewidth}\epsfig{file=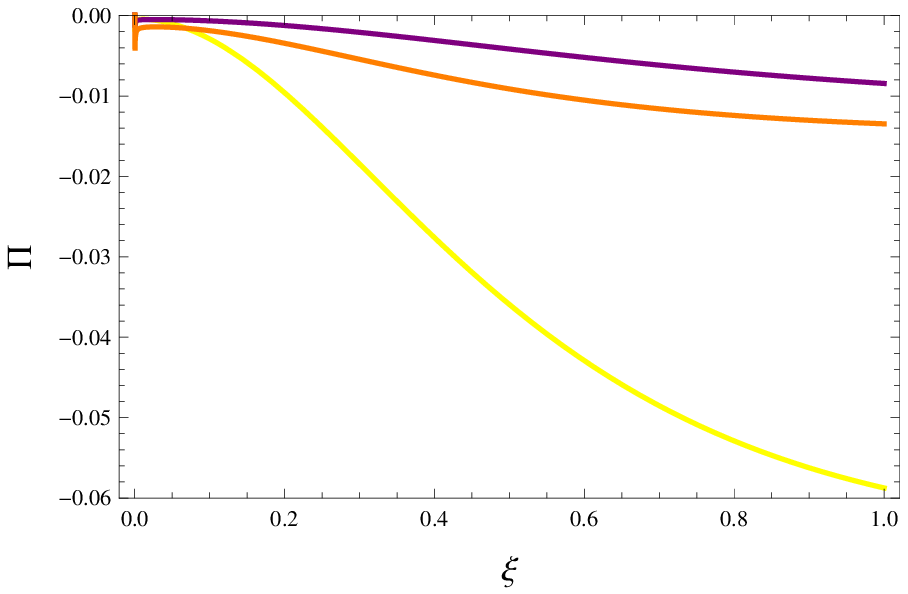,width=0.5\linewidth}
\caption{Plots of anisotropy versus $\xi$ for $n=3$, $\rho_o=5$,
$\zeta=7$, $\beta=0.001$.}
\end{figure}

For graphical analysis, we choose $n=3,~\rho_o=5,~ \zeta=7$ and
$\beta=0.0010$. Figures \textbf{1-3} show the behavior of
dimensionless energy density, mass function and anisotropy,
respectively. It can be seen from Figure \textbf{1} that $\Omega$ is
a decreasing function for smaller as well as greater values of
$\omega$. Moreover, the mass function has an inverse relationship
with $\omega$ while it varies directly with $\xi$. Further, we note
an increment in the anisotropy as $\omega$ rises from 1.3 to 1.7. In
contrast, the anisotropy within the system is negative when $\omega$
increases from 0.01 to 0.05. Thus, the physically acceptable
solution is associated with $\omega=1.3,~1.5,~ 1.7$.

\section{Conclusions}

In this paper, we have studied the impact of charge on the
complexity of a static sphere within the framework of EMSG. In this
respect, we have constructed the EMSG fields equations for a static
sphere by considering an anisotropic fluid distribution in the
presence of electromagnetic field. We have formulated the mass
functions $\emph{m}$ and $\emph{m}_{Tol}$ by utilizing the
definitions given by Misner-Sharp and Tolman, respectively. Their
link with Weyl tensor and matter variables has also been discussed.
We have also formulated the TOV equation in the context of EMSG. The
complexity factor has been obtained by formulating the structure
scalars through the orthogonal decomposition of the curvature
tensor. The scalar $\mathcal{Y}_{TF}$ accommodated the inhomogeneous
energy density, charge, anisotropic pressure and dark source terms
of $f(R,\textbf{T}^{2})$ gravity. Moreover, we have obtained the
expression of Tolman mass in terms of this strucuture scalar in the
presence of charge. Consequently, we have chosen $\mathcal{Y}_{TF}$
as a complexity factor. We have found that the addition of charge
decreases the complexity of stellar structure.

We have formulated the disappearing complexity condition by setting
$\mathcal{Y}_{TF}=0$. The vanishing complexity condition provides an
extra constraint which assists in obtaining the solution of field
equations by reducing the degrees of freedom. In this respect, we
have determined the complexity factor for a specific model,
$f\left(R,\textbf{T}^{2}\right)=R+\zeta \textbf{T}^{2}$. In GR, the
self-gravitating spherical system has zero complexity if it is
isotropic and homogeneous. However, in our work, an isotropic and
homogeneous system does not correspond to zero complexity which
implies the impact of dark source terms. Zero complexity is obtained
for
\begin{eqnarray}\nonumber
&&\int^{r}_0
\tilde{r}^{3}\left\{\left(\frac{1}{f_R}\right)^{'}\left(\rho+\varphi
+\varphi_{00}- \frac{\emph{q}^{2}}{8\pi r^{4}}\right)
+\left(\frac{1}{f_R}\right)\Big(\varphi+\varphi_{00}-
\frac{\emph{q}^{2}}{8\pi
r^{4}}\Big)^{'}\right\}\emph{d}\tilde{r}\\\nonumber
&&+\frac{1}{f_R}\left(\frac{\left(\varphi_{22}-\varphi_{11}\right)}{2}+\frac{\emph{q}^{2}}{4\pi
r^{4}}\right)=0.
\end{eqnarray}
Hence, the presence of additional matter terms of
$f(R,\textbf{T}^{2})$ gravity has enhanced the complexity of stellar
structure.

Finally, we have developed two solutions of EMSG field equations by
assuming $\emph{q}(r)=\beta r^{3}$. Firstly, to investigate the
features of stellar objects, we have assumed the energy density of
compact structure suggested by Gokhroo and Mehra. Secondly, we have
applied the polytropic equation of state and constructed a system of
dimensionless equations by introducing some new variables. This
system contained the dimensionless TOV equation, mass and
disappearing complexity condition. We have graphically analyzed the
numerical solution of this system by varying the parameter $\omega$.
The system has positive energy density for smaller as well as larger
values of $\omega$. However, the anisotropy corresponding to smaller
values of $\omega$ is negative. Thus, we have concluded that the
behavior of the considered system is physically acceptable for
$\omega=1.3,~1.5,~1.7$. It is important to mention here that results
of GR \cite{6} can be retrieved corresponding to $\zeta=0$.\\\\
\textbf{Data availability:} No new data were generated or analyzed
in support of this research.

\end{document}